# Fully symmetric controllable integrated three-resonator photonic molecule

JIAWEI WANG,[1] KAIKAI LIU,[1] QIANCHENG ZHAO,[1] ANDREI ISICHENKO,[1] RYAN Q. RUDY,[2] DANIEL J. BLUMENTHAL[1,*]

[1]*Department of Electrical and Computer Engineering, University of California Santa Barbara, Santa Barbara, CA 93106, USA*
[2]*U.S. Army Research Laboratory, Adelphi, Maryland 20783, USA*
*\* danb@ucsb.edu*

**Abstract:** Photonic molecules can be used to realize complex optical energy states and modes, analogous to those found in molecules, with properties useful for applications like spectral engineering, analog computation, many body physics simulations, and quantum optics. It is desirable to implement photonic molecules using high quality factor (Q) photonic integrated ring resonators due to their narrow atom-like spectral resonance, tunability, and the ability to scale the number of resonators on a photonic circuit. However, in order to take full advantage of molecule spectral complexity and tuning degree of freedom, resonator structures should have full symmetry in terms of inter-resonator coupling and resonator-waveguide coupling as well as independent resonance tuning, and low power dissipation operation, in a scalable integration platform. To date, photonic molecule symmetry has been limited to dual- and triple-cavity geometries coupled to single- or dual-busses, and resonance tuning limited to dual resonator molecules. In this paper, we demonstrate a three-resonator photonic molecule, consisting of symmetrically coupled 8.11 million intrinsic Q silicon nitride rings, where each ring is coupled to the other two rings as well as to its own independent bus. The resonance of each ring, and that of the collective molecule, is controlled using low power dissipation, monolithically integrated thin-film lead zirconate titanate (PZT) actuators that are integrated with the ultra-low loss silicon nitride resonators. This performance is achieved without undercut waveguides, yielding the highest Q to date for a PZT controlled resonator. This advance leads to full control of complex photonic molecule resonance spectra and splitting in a wafer-scale integration platform. The resulting six tunable supermodes can be fully controlled, including degeneracy, location and splitting as well as designed by a model that can accurately predict the energy modes and transmission spectrum and tunable resonance splitting. This symmetrically coupled three-resonator molecule opens the door to applications such as optical circulators, dispersion engineering, nonlinear frequency synthesis, analog computation and quantum photonic circuits and physics simulations that involve scalable number of molecules on chip.

## 1. Introduction

Photonic molecules (PM) are systems of coupled atom-like optical resonators that produce rich quantized energy states and supermodes with behavior analogous to atoms and molecules [1,2]. These characteristics enable precision control of light and light-matter interactions including complex dispersion engineering [3–7] and nonlinear energy-level transitions [8], and applications including many body physics simulations [9] and quantum optical phenomena [10]. The field of photonic molecules has evolved from the first demonstration of simple 2-coupled semiconductor optical cavities [2] to today's dual cavity, controllable, waveguide resonators capable of nonlinear 2-level transitions [8], symmetry breaking optical isolation [11] and soliton optical frequency comb generation [12]. Several dynamic tuning methods have been demonstrated including thermal [12], electrooptic [8], mechanical [13,14], optical [15] and acousto-optic [16]. However, to date, only electrooptic tuning has been shown to yield good independent tuning and low power dissipation [17]. The next stage of photonic molecule development requires scaling in terms of number of resonators per molecule, number of molecules, complexity of energy modes, controllability of the resonance linewidth and splitting, and power dissipation. Achieving these characteristics requires fabrication techniques capable of achieving high Q tunable coupled resonators coupled to low loss buses while maintaining low-power consumption. Advances in photonic molecule scaling, control, and integration, will enable applications including optical circulators, dispersion engineering, nonlinear frequency synthesis, analog optical computation and quantum photonic circuits and physics simulations.

Progress in photonic molecule integration has been limited to coupling between a small number (two to three) of microring resonators [18–20], microdisks [21–23] and photonic crystals [24,25]. More recent dual resonator advances include low power, electrooptically tunable molecules for soliton optical frequency comb generation [12] and a dual resonator optical isolator without tuning [11]. These demonstrations have had limited tunability, symmetry, and bus coupling, where molecules with more than two resonators are limited to chain-linked two-two coupling [26,27] or serially coupling [19] and therefore lacking full symmetry. Increasing the symmetry between resonators also requires providing access to each with its own bus waveguide, which to date has been limited to two buses for two- and three-resonator PMs. Enabling the DC bias resonance tuning of each resonator independently will open up new design flexibility, controllability, predictability and reliability for the field of photonic molecules. Tuning that can extend from MHz to GHz range is desirable for many photonic molecule applications. Low power actuators with good static (DC) tuning capabilities, including piezoelectric materials such as PZT and aluminum nitride (AlN) are highly attractive [28–30]. Additionally, these techniques should be compatible with fully planar, ultra-low loss platforms, without requiring complex under-etching processes. AlN has been used for tuning and control, however, the stress optical effect of AlN is relatively weak compared to PZT at frequencies under 1 GHz. As the number of resonators is increased, with each ring coupled to an independent bus, resonance and splitting control must be actuated using low power techniques, making the stress optic effect a better choice than thermal tuning. PZT actuation achieves low-power, strong DC tuning, and moderate AC modulation [31].

We demonstrate, for the first time to the best of our knowledge, a photonic molecule composed of three PZT tunable, ultra-low loss silicon nitride [32] high-Q microring resonators, in a fully symmetric geometry where each ring is coupled to all other rings and independent bus waveguide. This symmetric structure enables tuning of 6 supermodes that can be split further into programmable supermodes, that are coupled to the independent buses, yielding more degrees of tuning freedom than single- or dual-bus three resonator photonic molecules. Additionally, our tuning structure is realized using monolithically integrated, low power dissipation PZT actuators, that are compatible with our ultra-low loss silicon nitride platform without undercutting [32]. The PM is composed of three symmetrically coupled 580 μm radius silicon nitride ring resonators, with a measured 43 MHz full-width-at-half-maximum (FWHM) resonance width and a 48 GHz free spectral range (FSR). An 8.11 million intrinsic Q is measured for each ring operating at 1550 nm, which is the highest Q reported to date for a PZT controlled photonic integrated ring resonator. The resonance can be tuned by 2.5 GHz with a 15 V DC applied voltage, with a linear tuning sensitivity of 0.16 GHz/V and a low power consumption of 90 nW. We use a coupled photonic molecule (CPM) model to accurately model and predict the resonance tuning and splitting behavior, showing excellent agreement with experimental measurements. The CPM model provides a robust tool for photonic molecule device design and enables control of photonic molecules for integrated photonic applications including photonic molecule combs [12] and photonic molecule quantum photonics [10].

## 2. Results

### *2.1 Fully Symmetric Molecule Design and Fabrication*

The photonic molecule consists of three microring resonators symmetrically coupled to each other and to three independent bus waveguides, as shown in Fig. 1(a). This results in a PM that is fully symmetric to the three-axes angled at 120 degrees offset as indicated by the dotted lines in Fig. 1(a). The waveguide structures are fabricated using an ultra-low-loss silicon nitride platform [32,33] with a waveguide core dimension of 175 nm height and 2.2 μm width. The waveguide and actuator structure are shown in Fig. 1(b) (further fabrication details are provided in the Methods Section) without requiring undercut or released tuning structures and waveguides. An optical microscope image of the fabricated device with monolithically integrated PZT actuators is shown in Fig. 1(c).

The PZT actuator is designed with a 2 μm offset gap from the waveguide in order to reduce the optical mode loss while maintaining a relatively large strain-optic effect. The bus coupled resonator design is a 580 μm radius and 1.5 μm ring-bus coupling gap. We measure the resonance FWHM to be 43.23 MHz at 1550 nm, as shown in Fig. 1(d), using a radio frequency calibrated unbalanced Mach-Zehnder interferometer (MZI) [34,35]. These measurements correspond to a loaded Q factor of $4.48 \times 10^6$ and an intrinsic Q factor of $8.11 \times 10^6$, corresponding to a propagation loss of 3.3 dB/m. As a comparison, the resonator of the same geometry but without PZT actuator has a resonance FWHM of 36.60 MHz, as shown in Fig. 1(e), corresponding to 5.28 million loaded Q factor, 8.37 million intrinsic Q factor and propagation loss of 3.19 dB/m. By comparing the two measurements, the intrinsic Qs are similar, and the loss increase due to the PZT and its metal electrodes is only 3.4%, which demonstrates that the monolithically integrated PZT actuators are compatible with our ultra-low-loss silicon nitride platform.

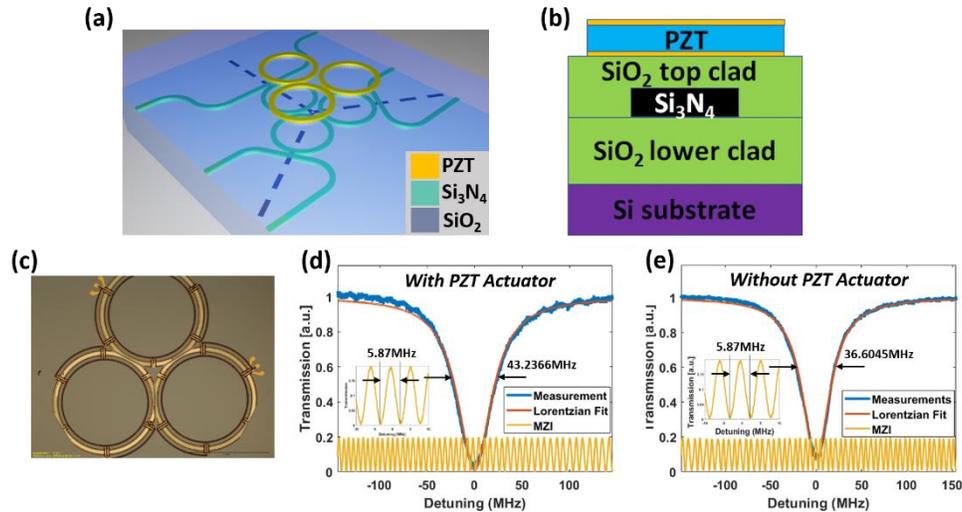

Fig. 1. (a) 3D schematic illustration of the symmetrically coupled photonic molecule. The three ring resonators, each of the same radii, are symmetrically positioned relative to the three dotted axes angled at 120 degrees from each other. All inter-ring gaps are equal, and all ring-bus waveguide gaps are also equal. (b) The tunable waveguide cross-section geometry. (c) Optical micrograph of a fabricated device with monolithically integrated lead zirconate titanate (PZT) actuators. (d) The quality factor (Q) measurement of a resonator with R = 580 μm with PZT actuator. The Mach-Zehnder interferometer (MZI) fringe pattern (yellow trace) with an FSR of 5.87 MHz works as a frequency ruler to measure the full-width-at-half-maximum (FWHM) of the resonance (blue trace). The loaded Q is measured to be 4.48 million and intrinsic Q is derived to be 8.11 million at 1550 nm. (e) The Q measurement of the resonator of same dimension but without PZT actuator deposition. The loaded Q is measured to be 5.28 million and intrinsic Q is derived to be 8.37 million.

## 2.2 Photonic Molecule Model

A coupled mode theory (CMT) model [36] is used to determine the supermode behavior, validate the measured resonance structure, and enable programming of desired spectral properties. To date, CMT modeling has been limited to three resonators that are serially coupled [19] or side-by-side coupled in a linear chain structure [27]. We employ a matrix approach to model the fully symmetric three ring photonic molecule that includes the waveguide loss, bus-ring and ring-ring coupling coefficients, and effective mode index. Further details of the modeling and calculation are given in the supplementary section.

Each resonator supports two modes (shown as $A_1$ through $C_6$ in Fig. 2(a), defining the clockwise (CW) and the counterclockwise (CCW) coupled propagating fields. Each photonic atom (resonator) can be considered to have two split states with a bonding orbital and an antibonding orbital as in the hydrogen molecule model [37], where these states coherently add to generate supermodes. The transfer matrix yields six eigenvalues that represent the six energy supermodes. We study three cases in order to understand mode splitting as depicted in Fig. 2 (b)-(d). When all resonators and waveguides and coupling parameters are identical, the molecule is perfectly symmetric, leading to a degenerate solution with 4 supermodes as shown in the transmission spectrum in Fig. 2(b), with the rightmost supermode as the highest energy level and the leftmost supermode the lowest energy level. The four remaining supermodes come in two degenerate pairs [38]. When symmetry is broken, for example if the coupling strengths κ between the resonators are non-uniformly changed, the middle degenerate pairs are split and mode degeneracy is lifted, as shown in Fig. 2(c). Fabrication variations among resonators and coupling coefficients move the system away from full symmetry and as such, the supermodes evolve as shown in the experimental and modeled energy spectra in Fig. 2(d). The blue line is a measured spectrum, and the orange line is our CPM model fitting that incorporates measured cavity loss γ and inter-ring coupling coefficients κ. The excellent agreement between the modeled (fit) and measured transmission spectrum verifies the accuracy of our CPM model. We also simulate the three cases using Lumerical INTERCONNECT®, which agree with both the CPM modeling and the measurements. Being able to design and predict the mode energy splitting is important for applying photonic molecules.

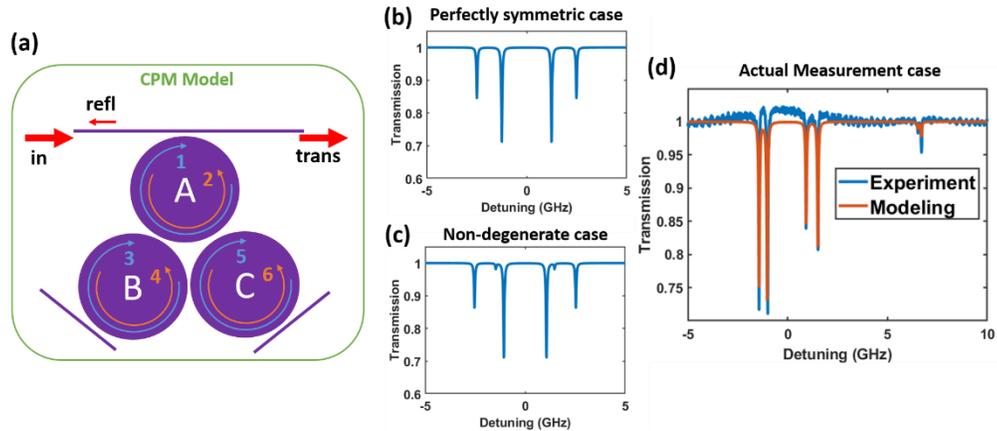

Fig. 2. (a) Coupled photonic molecule (CPM) model, where each photonic atom has two split states as a result of the clockwise and counterclockwise propagating fields and mutual coupling. (b) Simulated transmission spectrum of the aligned case where all the resonators and coupling coefficients are identical. The middle four supermodes are two degenerate pairs. (c) The non-degenerate case occurs when the coupling strength between the photonic atoms are unbalanced and symmetry is broken, and the middle degenerate supermodes pairs split. (d) Measurement and model fitting when the measured losses and coupling coefficients are utilized in CPM.

## 2.3 Resonance and Supermode Tuning

We demonstrate tunable PM supermodes by controlling the DC bias of the PZT actuator for each ring resonator. The DC bias applies electric field to the PZT film and causes it to strain, that in turn induces effective refractive index change of the optical mode of the waveguide beneath the actuator via the stress-optic effect [39]. Independent DC bias for each ring resonator enables full control over supermode splitting and frequency location. An example is shown in Fig.3 (a)-(c). The colored lines indicate the measured transmission spectra as the DC bias voltages applied to the PZT actuators are adjusted from 0 V to 15 V on each ring. The modeled behavior, given by the dashed lines, are in good agreement with the measurements, yielding an important tool for designing, predicted, and accurately controlling the supermodes. As shown in Fig.3 (d)-(f), the resonances can be detuned by 2.5 GHz at 15 V bias voltage, with an average linear tuning coefficient of 0.16±0.03 GHz/V, that is 10 times larger compared to that of AlN actuation operating at 150 V, without released structure [30]. Fine tuning adjustment of the transmission spectra to overcome fabrication variations is demonstrated in the expanded-view panel of Fig.3 (c), which is close to the case described in Fig.2 (c). By tuning two or three of the resonators at the same time the symmetric degenerate case can be reached, providing a way to calibrate and balance the photonic molecule.

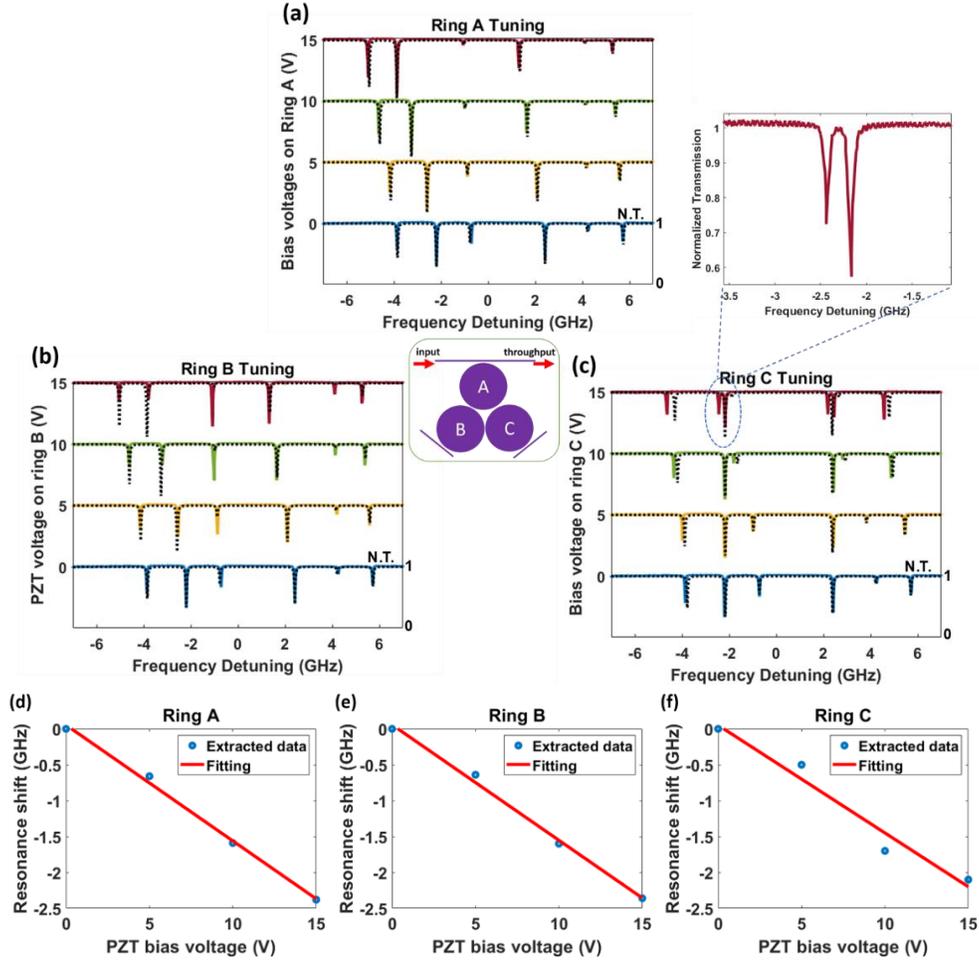

Fig. 3. (a)-(c): DC biasing of rings A, B and C. The colored lines indicate the measured transmission spectrums as DC bias is adjusted from 0 to 15V and the dashed black lines are the coupled photonic molecule model fitting; (d)-(f): DC tuning curves for each resonator with an average tuning sensitivity of 0.161 GHz/V; N.T., normalized transmission.

The PZT material has high resistivity resulting in a very low leakage current and power dissipation. We measure the leakage current to be about 6 nA at 15 V bias, corresponding to 90 nW electrical power consumption measured for each actuator, orders of magnitude lower than a thermal heater (30 mW) [40] and comparable to that of AlN actuation (300 nW at 150V) [30]. The capacitance of the PZT actuator is measured to be around 600 pF, corresponding to a stored energy of 67 nJ at 15 V. Table 1 compares our work to a recently published AlN actuated resonator [30]. This table highlights the advantage that the PZT actuator offers a nearly 10 times larger tuning effect while maintaining low power consumption, as compared to AlN.

Table 1. Comparison between PZT and AlN actuation and power consumption

|  | PZT actuator in this work | State-of-the-art AlN actuator [30] |
|---|---|---|
| Tuning Coefficient | 160 MHz V$^{-1}$ | 15.7 MHz V$^{-1}$ |
| Power Consumption | 90 nW | 300 nW |

## 3. Discussion

In this paper, we have demonstrated the first, to the best of our knowledge, fully symmetric photonic molecule composed of three mutually and symmetrically coupled microring resonators. The molecule consists of 580 μm radius silicon nitride resonators with measured intrinsic Q of $8.11\times10^6$, which is the highest Q reported to date for a PZT controlled photonic integrated ring resonator, to the best of our knowledge. Each resonator can be controlled using monolithically integrated 90 nW low-power dissipation PZT actuation, with a linear tuning coefficient of 0.16 GHz/V. The devices are fabricated with a CMOS-compatible process without requiring complex processes like under-cut or released and suspended structures.

A matrix coupled photonic molecule model is developed to accurately predict and simulate the behavior of the six supermodes. Supermode evolution and resonance splitting are discussed, modeled, and measured, with the predicted and measured responses in good agreement. The photonic molecule can be tuned into the degenerate and the non-degenerate states by adjusting the bias voltages, allowing compensation for fabrication variations between the rings and coupling gaps. By changing coupling strengths between the resonators with the piezo actuation, the peak narrowing and extinction ratio can be controlled.

The independent tuning and control using PZT actuators provides a programmable and scalable PM platform that increases the design flexibility, particularly for DC bias and calibration. Future work includes AC modulated PZT actuators, with responses expected out to several hundred MHz to a GHz. These actuators will lead to new functions like optical isolation via spatio-temporal modulation [41] as well as sideband modulation for locking in atomic, molecular and optical (AMO) physics applications including atomic clocks [42,43], atomic and molecular spectroscopy [44–46], and quantum sensing [42,47,48]. The dynamic tuning can provide new active control photonic component solutions, for example to lock a stimulated Brillouin scattering (SBS) resonator to the pump laser [34], provide feedback actuation inside the SBS laser cavity itself, and manipulate the soliton repetition rate of a microcomb [30]. It should be noted that AlN has extremely good high frequency response and has been successfully demonstrated in a wafer-scale photonic platform [29] and integration these two actuators together to provide strong DC and broadband AC response is a subject of future work. The multiple discrete levels of the photonic molecule can be utilized as a signal splitter or a wavelength shifter with proper microwave tuning and cavity design for unidirectional flow [49,50]. These results provide a path towards and integration platform that enables study of multiple coupled microcavities with varying configurations for applications like multichannel high-order filters [20], topological photonics [5], quantum signal processing [10], analog computation, and many body physics simulations [9]. The low-power consumption is critical to enable large-scale integration of coupled photonic atoms and complex photonic molecules on a chip, and suited for heterogeneous integration with semiconductor lasers [51].

## Methods.

### Fabrication Process

The fabrication begins with a 1 mm thick, 4-inch diameter silicon substrate with 15 μm of thermally grown silica on top as lower cladding. Then 175 nm of stoichiometric silicon nitride film is deposited by low-pressure chemical vapor deposition (LPCVD). The waveguide is patterned using a photoresist mask by 248 nm DUV stepper lithography and inductively coupled plasma etch with $CF_4/CHF_3/O_2$ gas. Next, 6 μm thick of $SiO_2$ is deposited on top of the waveguide as upper cladding by tetraethylorthosilicate (TEOS)-based plasma enhanced chemical vapor deposition (PECVD). After upper cladding deposition, the wafer is chemical-mechanical polished (CMP) to planarize the surface for the piezo actuator deposition. The actuator stack consists of a sputtered 40 nm thick TiO2 adhesion layer, a sputtered 150 nm thick Pt bottom electrode, a 500 μm thick layer of PZT (52/48 Zr/Ti ratio) is deposited via chemical solution deposition and the stack is capped with a sputtered 100 nm thick Pt top electrode. The PZT and Pt electrodes are patterned by argon ion milling. With the actuator patterned, electrical traces are evaporated and patterned through lift-off and consist of a Cr / Pt / Au stack with thicknesses of 20 nm / 20 nm / 730 nm respectively. To reduce resistivity and minimize thick gold coverage on the electrodes a 10 μm thick copper layer is electroplated using photoresist molds and a sputtered copper seed layer. The photoresist molds and copper seed layer are removed with solvents to release the device.

### Quality factor measurements and calculation

A tunable laser (Velocity TLB-6730) is controlled by an external waveform generator for frequency detuning. The optical power is split into two, one fraction goes to the resonator and one fraction goes to an unbalanced Mach-Zehnder interferometer (MZI) which has a calibrated free spectral range of 5.87 MHz. The transmitted power from the device under test and the MZI are monitored on a synchronized oscilloscope. The MZI fringe patterns provide a radiofrequency calibrated frequency reference for accurate evaluation of resonance. The full width at half maximum (FWHM) of the resonator is measured by fitting the resonance spectrum to a Lorentzian curve and counting the corresponding MZI peaks.

### Characterization of resonance tuning and power consumption

The transmission spectrum of the throughput of the three resonators photonic molecule is measured with a tunable laser (Agilent 81680A) and an optical power sensor (Agilent 81634B) embedded with the Agilent 8164A Lightwave Measurement System. The DC voltages are applied to the electrical pads of the PZT actuators by programmable DC power supplies (Keithley 2602A SourceMeter). The leakage current and the consumed electrical power is measured by a precision source and measure unit (Keysight B2902A) with 100 fA measurement resolution.

## Funding.
This work was supported by the National Science Foundation under EAGER Grant No. 1745612 and the Advanced Research Projects Agency-Energy (ARPAE), U.S. Department of Energy, under Award Number DE-AR0001042. The views, opinions, and/or findings expressed are those of the authors and should not be interpreted as representing the official views of the U.S. Government.

**Acknowledgment.**
The samples were fabricated in the UCSB NanoFabrication Facility and US Army Research Laboratory. The authors would like to thank Steven Isaacson of General Technical Services for his efforts in fabrication of the devices. We want to thank Dr. Michael Zervas and Dr. Davide Sacchetto for their help on CMP of the wafer at LiGenTec SA. A.I. acknowledges the support from the National Defense Science and Engineering Graduate (NDSEG) Fellowship Program.

**Disclosures.**
The authors declare no conflicts of interest.

**Data Availability.**
The data that support the plots within the paper and other findings of this study are available from the corresponding author upon reasonable request.

**Supplemental document.**
See Supplement 1 for supporting content.

# Fully symmetric controllable integrated three-resonator photonic molecule: supplemental document

This document provides supplementary information to "Fully symmetric controllable integrated three-resonator photonic molecule". In section 1, we present the matrix modeling of the coupled photonic molecule. In section 2, we show the simulation of the symmetrically coupled three resonators photonic molecule done in Lumerical INTERCONNECT. In section 3, we explain the math to calculate and extract the intrinsic and loaded Q factors and the propagation loss.

## 1. Coupled Photonic Molecule Modeling

Each photonic molecule has two degenerate modes at each resonant frequency, the clockwise (CW) and the counterclockwise (CCW) propagating, as shown in Fig. S1(a). This can be understood by bringing the analog of the hydrogen model [1] where each hydrogen atom has two split states due to the bonding and antibonding orbitals as shown in Fig. S1(b). Therefore, there are six modes in a three mutually coupled resonator system. The coupling between these modes is determined by the gap between the resonators. The coupled photonic molecule (CPM) model is formulated in a matrix approach, where the diagonal elements specify the resonance frequency and loss rate for each mode and the non-diagonal elements specify the coupling rates between the modes.

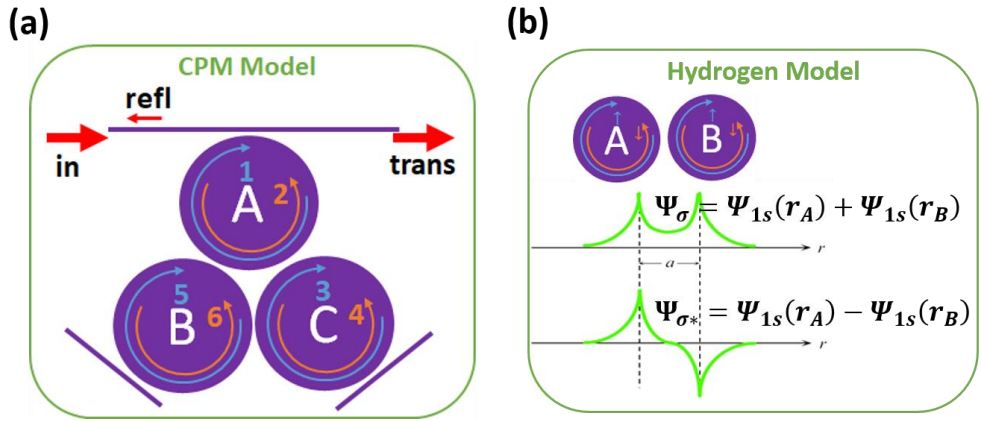

Fig. S1. (a) Coupled photonic molecule model and (b) the comparison with hydrogen model where the bonding and antibonding orbital states can be analogous to the clockwise (CW) and counterclockwise (CCW) propagating mode.

The general transfer matrix therefore can be developed as:

$$\frac{d}{dt}\begin{bmatrix}a_1\\a_2\\a_3\\a_4\\a_5\\a_6\end{bmatrix}=\begin{bmatrix}-\frac{\gamma}{2}+i\omega_A & 0 & 0 & i\kappa_{12} & 0 & i\kappa_{13}\\0 & -\frac{\gamma}{2}+i\omega_A & i\kappa_{12} & 0 & i\kappa_{13} & 0\\0 & i\kappa_{12} & -\frac{\gamma}{2}+i\omega_B & 0 & 0 & i\kappa_{23}\\i\kappa_{12} & 0 & 0 & -\frac{\gamma}{2}+i\omega_B & i\kappa_{23} & 0\\0 & i\kappa_{13} & 0 & i\kappa_{23} & -\frac{\gamma}{2}+i\omega_C & 0\\i\kappa_{13} & 0 & i\kappa_{23} & 0 & 0 & -\frac{\gamma}{2}+i\omega_C\end{bmatrix}\begin{bmatrix}a_1\\a_2\\a_3\\a_4\\a_5\\a_6\end{bmatrix}+\begin{bmatrix}1\\0\\0\\0\\0\\0\end{bmatrix}i\sqrt{\gamma_c}s_{in}$$

(S1)

where $a_i$ represents the modal amplitudes of the supermodes, $\omega_i$ is the resonant frequency of individual photonic atom, $\kappa_{ij}$ is the coupling rate between the ith and jth modes, $\gamma$ is the loss rate, $\gamma_c$ represents the bus-ring coupling rate and $s_{in}$ is the optical input in the bus waveguide.

Thus, the transmission from the input port to the throughput of the PM can be written as:

$$T = \left|1 + i\sqrt{\gamma_c}a_1\right|^2$$

(S2)

## 2. Coupled Photonic Molecule Simulation

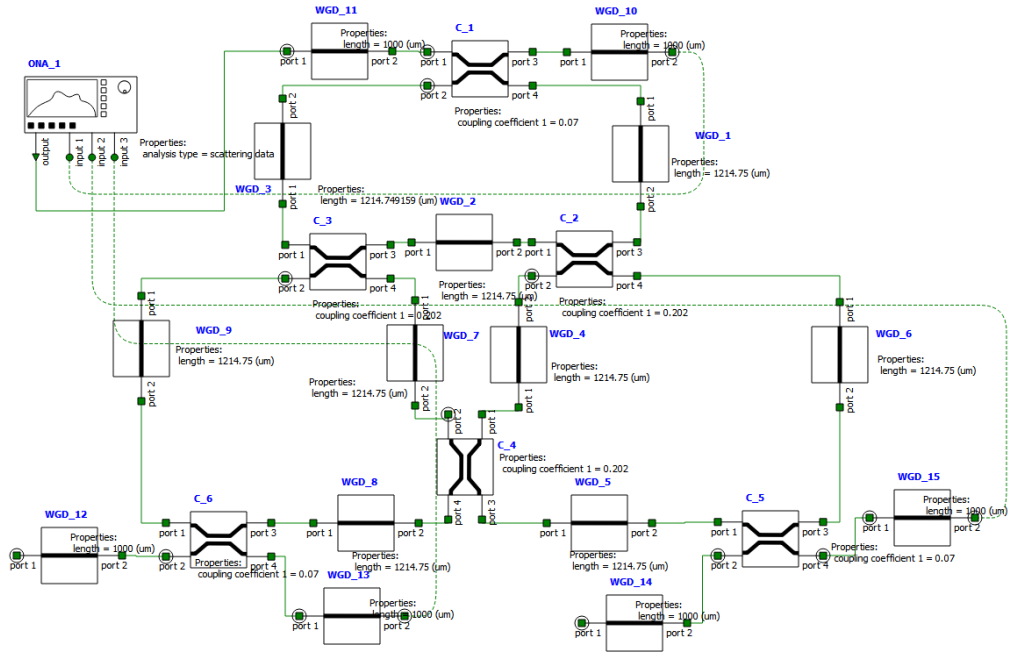

Fig. S2. Lumerical INTERCONNECT schematic diagram

As shown in Fig. S2, the PM is simulated in Lumerical INTERCONNECT® using the waveguide and the coupler components to realize the three symmetrically coupled resonator system by incorporating device waveguide length, waveguide losses, effective mode index and coupling coefficients. The comparison between the CPM modeling and the device simulation is shown in Fig. S3, and the accuracy between the modeling prediction, simulation verification and measurements can be confirmed.

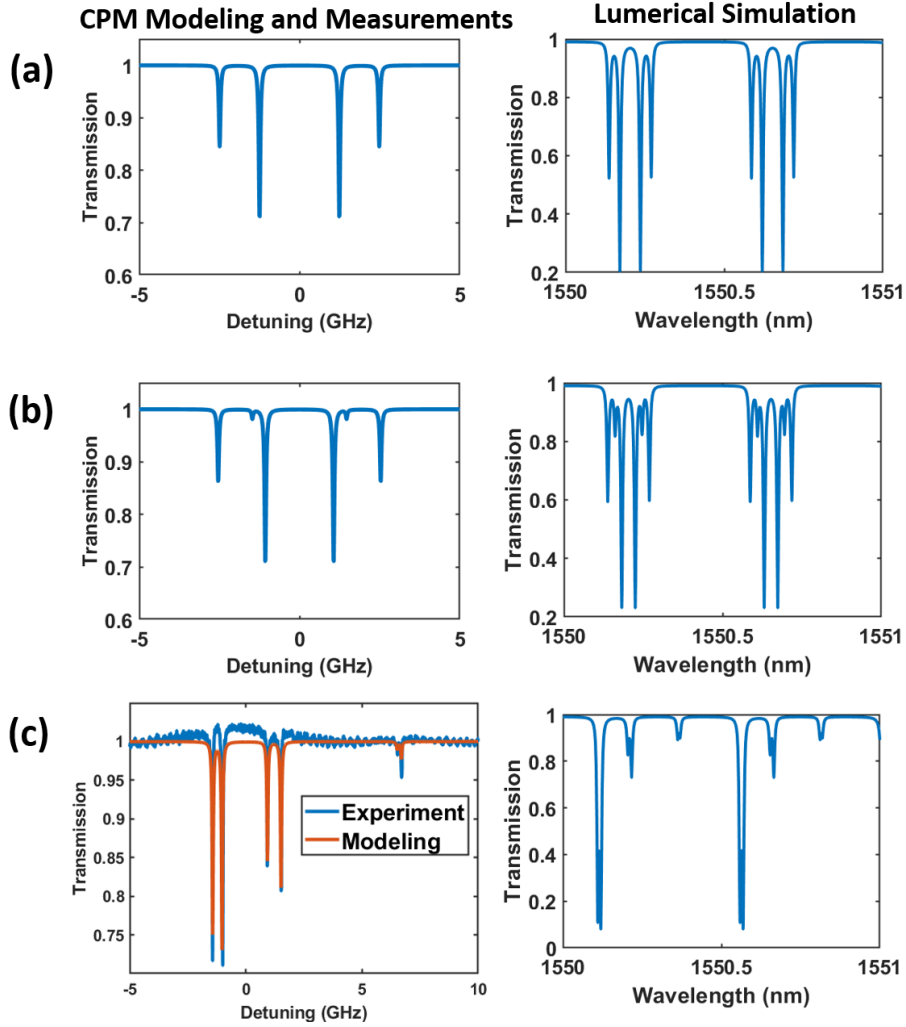

Fig. S3. Comparison between the coupled photonic molecule (CPM) modeling (left side) and Lumerical simulation for two free-spectral-range (FSR) (right side).

## 3. Quality factor calculation and loss extraction

The propagation loss of the waveguide is extracted based on the following equation [2,3],

$$Q_{Load} = \frac{\lambda_{res}}{FWHM} = \frac{\pi n_g L \sqrt{ra}}{\lambda_{res}(1-ra)} \tag{S3}$$

where $Q_{Load}$ is the loaded quality factor measured from with a radio frequency calibrated unbalanced Mach-Zehnder interferometer (MZI) [3]. $n_g$ is the group index of the waveguide,

$L = 2\pi R$ is the perimeter of the ring resonator, $\lambda_{res}$ is the resonant wavelength, $r = \sqrt{1 - \kappa^2}$ is the self-coupling coefficient and $\kappa^2$ is the power coupling coefficient, $a$ is the single-pass amplitude transmission and is related to the power attenuation coefficient $\alpha$ as $a^2 = exp(-\alpha L)$. The intrinsic Q of the resonator can be calculated with the extraction of waveguide propagation loss $\alpha$ using the following equation [4].

$$Q_{int} = \frac{2\pi n_g}{\lambda_{res} \alpha} \qquad (S4)$$